\documentclass[journal]{IEEEtran}
\usepackage{rotating}
\usepackage{amsmath}
\usepackage{graphicx}
\usepackage{subfigure}
\usepackage{algpseudocode}
\usepackage{algorithm}
\usepackage{algorithmicx}
\usepackage{algpseudocode}
\usepackage{amssymb}
\usepackage{float}
 \usepackage{verbatim}
\usepackage{multirow}
\usepackage{makecell}
\usepackage{hyperref}
\usepackage{graphics} 
\usepackage{diagbox}
\usepackage{color}
\usepackage{multirow,booktabs}
\usepackage{adjustbox}

\usepackage[table]{xcolor} 
\usepackage{cleveref}

\usepackage[numbers,sort&compress]{natbib}

\begin{document}

\title{WK-Pnet: FM-Based Positioning via Wavelet Packet Decomposition and Knowledge Distillation}

\author{Shilian~Zheng,
        Quan~Lin,
        Peihan~Qi,
        Luxin~Zhang,
        Xinjiang~Qiu,
        Zhijin~Zhao,
        and Xiaoniu~Yang

\thanks{This work was supported in part by the National Natural Science Foundation of China under Grant 62171334. \textit{(Corresponding author: Shilian Zheng and Peihan Qi.)}}

\thanks{S. Zheng is with the College of Communication Engineering, Hangzhou Dianzi University, Hangzhou 310018, China. He  is also with National Key Laboratory of Electromagnetic Space Security, Jiaxing 314033, China (e-mail: lianshizheng@126.com).}


\thanks{Q. Lin, X. Qiu, and Z. Zhao are with the College of Communication Engineering, Hangzhou Dianzi University, Hangzhou 310018, China (e-mail: 232080162@hdu.edu.cn; 232080090@hdu.edu.cn; zhaozj03@hdu.edu.cn).}

\thanks{P. Qi is with the State Key Laboratory of Integrated Service Networks, Xidian University, Xi’an 710071, China (e-mail: phqi@xidian.edu.cn).}

\thanks{L. Zhang and X. Yang are with {{National Key Laboratory of Electromagnetic Space Security}},
Jiaxing 314033, China (e-mail: lxzhangMr@126.com; 
yxn2117@126.com).}

}

\maketitle

\begin{abstract}

Accurate and efficient positioning in complex environments is critical for applications where traditional satellite-based systems face limitations, such as indoors or urban canyons. This paper introduces WK-Pnet, an FM-based indoor positioning framework that combines wavelet packet decomposition (WPD) and knowledge distillation. WK-Pnet leverages WPD to extract rich time-frequency features from FM signals, which are then processed by a deep learning model for precise position estimation. To address computational demands, we employ knowledge distillation, transferring insights from a high-capacity model to a streamlined student model, achieving substantial reductions in complexity without sacrificing accuracy. Experimental results across diverse environments validate WK-Pnet’s superior positioning accuracy and lower computational requirements, making it a viable solution for positioning in real-time resource-constraint applications.


\end{abstract}

\begin{IEEEkeywords}
Positioning, deep learning, FM signals, wavelet packet decomposition, knowledge distillation.
\end{IEEEkeywords}

\section{Introduction}
With the rapid growth of the Internet of Things (IoT), the demand for accurate, reliable, and real-time positioning technologies has surged across various sectors, including autonomous driving \cite{laconte2021survey}, logistics tracking \cite{yayan2015comprehensive}, and health monitoring \cite{liu2021kalman}. Global Navigation Satellite Systems (GNSS), such as the Global Positioning System (GPS) \cite{el2002introduction}, perform well in signal-strong outdoor environments, offering high-precision positioning services. However, GNSS systems also exhibit notable limitations. Firstly, satellite positioning systems rely on precise signal synchronization, determining location by measuring the signal propagation time. Microsecond-level timing errors can lead to significant positioning deviations, thus requiring highly accurate time synchronization. Secondly, in complex environments (such as indoors or densely populated cities), the accuracy of these systems is substantially affected. In urban environments, signals are easily susceptible to electromagnetic interference, signal blockage, and obstructions from buildings, leading to signal attenuation, reduced positioning accuracy, and stability, which often fails to meet users' demands for accurate and real-time positioning services.

Opportunistic signal positioning has emerged as an auxiliary technology, offering a new solution to positioning challenges \cite{obeidat2021review,retscher2022indoor,asaad2022comprehensive}. Opportunistic signal positioning utilizes non-navigation radio frequency (RF) signals present in the environment for positioning purposes, such as WiFi signals \cite{feng2024wi}, ultra wideband (UWB) signals \cite{perez2020indoor}, digital video broadcasting-terrestrial (DVB-T) signals \cite{chen2016analysis}, mobile communication signals \cite{liu2023machine}, and visible light \cite{zhu2023positioning}. This approach does not require additional deployment of dedicated sensors or beacons, instead utilizing existing signal resources for a degree of flexibility and cost-effectiveness. Positioning devices can receive and process these environmental signals to analyze signal characteristics, achieving relatively accurate positioning.

Frequency modulation (FM) signals, as an opportunistic source of navigation, have demonstrated excellent positioning accuracy \cite{matic2010fm,popleteev2017indoor,cong2023practical}. FM signal transmission towers are usually located on high points within cities to ensure the signal can penetrate building exteriors and provide wide coverage. This setup reduces signal blind spots, effectively bridging indoor and outdoor positioning gaps. Additionally, the FM signal operates in the frequency range of 87 MHz to 108 MHz, with a wavelength of approximately 3 meters. Compared to other wireless signals like WiFi, FM signals have a longer wavelength, which helps reduce signal attenuation from obstacles in the environment and mitigates multipath effects. These properties allow FM signals to penetrate walls, floors, and other barriers, providing enhanced signal strength and stability in diverse environments \cite{obeidat2021review}.

FM signal-based positioning technologies can be primarily categorized into two types: signal propagation modeling and fingerprint recognition techniques. Most signal propagation modeling methods establishes a relationship between the received signal strength indicator (RSSI) and propagation distance, providing a reliable positioning approach through analysis of the correlation between signal strength and location \cite{yoon2015acmi,cong2020practical,chen2022learning,li2022fm,cong2024mems}. 
For example, Yoon et al. \cite{yoon2015acmi} utilized FM broadcast signals for positioning by constructing an FM signal propagation model, demonstrating the robustness and accuracy of FM signals for positioning. Cong et al. \cite{cong2020practical} proposed a floor positioning algorithm based on multi-feature motion mode recognition, combining MEMS inertial sensors with FM signals. Chen et al. \cite{chen2022learning} developed the RadioLoc system, which uses FM signals for vehicle positioning and incorporates modern machine learning techniques to process FM signals for efficient all-terrain positioning. Li et al.  \cite{li2022fm} introduced an online FM fingerprint database construction and calibration method based on the propagation model and pedestrian dead reckoning (PDR) for positioning. Cong et al. \cite{cong2024mems} designed an integrated MEMS-INS/GNSS navigation system with FM signal-aided distance estimation during GNSS outages, which uses support vector regression (SVR) for distance increment estimation and an extended kalman filter (EKF) for data fusion. Other widely used modeling methods include time of arrival (TOA), time difference of arrival (TDOA), and angle of arrival (AOA) \cite{shamaei2021joint,pan2022efficient,bottigliero2021low,huang2023range,wang2022cooperative,zou2023convergent}. However, these methods often require precise time synchronization and complex signal processing techniques to ensure accuracy, which can be challenging to achieve, especially with FM signals that lack precise timing characteristics.

Fingerprinting-based techniques, by contrast, provide a simpler alternative by streamlining the signal processing workflow and removing the need for precise time synchronization or angle measurements. 
For instance, Chen et al. \cite{chen2012fm} demonstrated the feasibility of FM signal fingerprinting for indoor positioning. Moghtadaiee and Dempster \cite{moghtadaiee2014indoor} developed an FM signal strength fingerprinting-based positioning system, comparing deterministic and probabilistic fingerprinting methods and proposing a new hybrid method to enhance positioning accuracy. Du et al. \cite{du2020kf} proposed an FM-based fingerprint positioning algorithm (KF-KNN), which integrates K-nearest neighbors (KNN) and a Kalman filter model to improve accuracy and adaptability to environmental conditions. Mukherjee et al. \cite{mukherjee2019losi} utilized FM signals and a software-defined radio (SDR) to capture the RSSI, enabling high-precision positioning across the continental United States through spectrum estimation. Popleteev  \cite{popleteev2019improving} conducted a year-long experiment demonstrating the robustness of FM signal features under various human activities and weather conditions and introduced new signal characteristics to improve positioning accuracy. However, these approaches primarily rely on RSSI as the fingerprint feature. By focusing solely on energy characteristics, they limit the ability to fully capture the rich information present in the original signal. Additionally, as these methods predominantly use traditional machine learning techniques, they may lack the capability to learn complex, deep features necessary for high positioning accuracy in challenging positioning tasks.


Deep neural networks (DNNs) have become a popular choice in positioning research due to their powerful data processing capabilities. The introduction of deep learning has brought breakthroughs to FM signal fingerprinting research. For instance, Lei et al. \cite{lei2022defloc} proposed an indoor vehicle positioning method based on FM signal fingerprint maps and deep learning, achieving over $90\%$ positioning accuracy even with a $60\%$ data loss rate. Zheng et al. \cite{zheng2024fm} introduced a deep learning-based positioning method, FM-Pnet, which represents FM signals using short-time fourier transform (STFT) and leverages deep learning for end-to-end positioning, achieving more accurate positioning than traditional methods. Effective signal preprocessing enables neural networks to learn signal features more efficiently and achieve more precise positioning. \textcolor{blue}{However, beyond STFT, time-frequency analysis techniques such as wavelet transform \cite{song2022intelligent} and empirical mode decomposition (EMD) \cite{zheng2024cooperative} also hold significant potential.} In particular, wavelet transform can decompose signals at multiple scales, making it especially effective for capturing local features of non-stationary signals, thus presenting a promising avenue for further exploration in positioning applications. Furthermore, current FM-Pnet models often employ complex neural networks, creating substantial computational demands that are not ideal for IoT environments, where lightweight and efficient processing is essential. Reducing the computational load and enhancing overall efficiency is a critical goal, as it allows these methods to be effectively deployed in resource-constrained IoT applications, supporting real-time and scalable positioning solutions.

To address these challenges, in this paper we propose a novel lightweight FM-based positioning method empowered by wavelet packet decomposition (WPD) and knowledge distillation, called WK-Pnet. WK-Pnet utilizes WPD to extract time-frequency features of FM signals, feeding these into a deep neural network for location estimation. Additionally, knowledge distillation is introduced to train a lightweight neural network model, which significantly reduces the computational load while maintaining positioning accuracy. The WK-Pnet method comprises two main phases: offline training and online inference. During the offline training stage, FM signals from various known locations are collected to establish a dataset and train the model parameters. In the online inference stage, the system receives FM signals from an actual location, feeds them into the pre-trained model, and estimates the current location based on the model’s output confidence. The primary contributions of this study are as follows:

\begin{itemize}
\item We propose WK-Pnet, an innovative FM signal-based positioning framework. At its core, WK-Pnet leverages WPD to extract the time-frequency features of FM signals. This process not only enhances signal representation but also provides rich feature input to the deep learning model, achieving higher positioning accuracy in both indoor and outdoor environments.
\item We adopt a knowledge distillation framework to achieve model lightweighting while ensuring positioning accuracy. Using knowledge distillation, we transfer knowledge learned by a complex model to a simpler model. This approach greatly reduces computational resource consumption, improves model processing speed, and maintains high positioning accuracy of WK-Pnet.
\item We conduct extensive experiments to evaluate the performance of WK-Pnet in various scenarios. In our experiments, \textcolor{blue}{we compare the impact of different time-frequency inputs (STFT, EMD, and WPD)} and analyze the teacher model fine-tuned from FM-Pnet. Additionally, we compare the performance of the knowledge distillation model, non-distilled model, and FM-Pnet. The experimental results demonstrate that WK-Pnet consistently achieves outstanding performance across all tests.

\item \textcolor{blue}{We analyze the complexity of the proposed method, including its time complexity (measured in floating-point operations, FLOPs) and space complexity (measured by the number of model parameters, Params). The results show that, compared to FM-Pnet, our proposed WK-Pnet reduces FLOPs by approximately $95.9\%$ and Params by about $99.3\%$ without sacrificing positioning accuracy.}

\item \textcolor{blue}{We evaluate the impact of different temporal-frequency inputs and models on inference latency on mobile edge devices and deploy them for testing on a smartphone. The experimental results show that, compared to FM-Pnet, WK-Pnet reduces the inference latency by $90.5\%$.}
\end{itemize}

The rest of the paper is organized as follows. In Sec. \ref{sec1}, we formulate the positioning problem. In Sec. \ref{sec2}, we introduce our proposed WK-Pnet. In Sec. \ref{sec3}, we compare the  performance of WK-Pnet with existing FM-Pnet. Finally, Sec. \ref{sec4} provides the concluding remarks.

\section{Problem Formulation}
\label{sec1}
 
 \begin{figure}[t]
    \centering 
    \includegraphics[width=8cm]{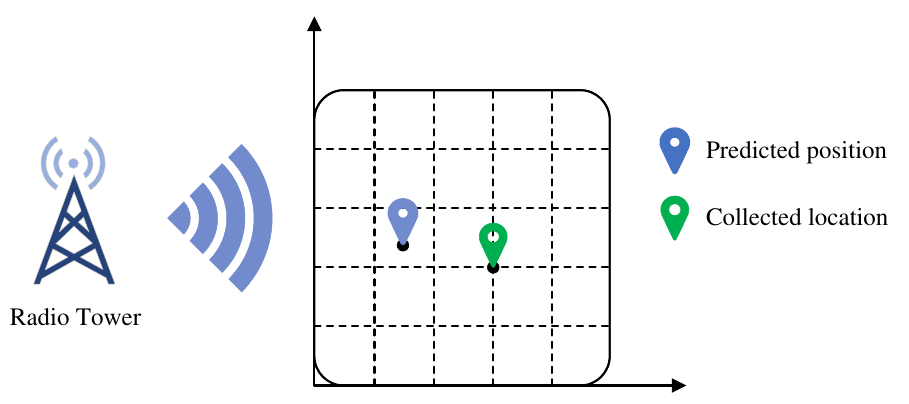}  
    \caption{Problem modeling for WK-Pnet Positioning.}
    \label{fig1}
\end{figure}

FM signal fingerprinting differs from distance-based positioning techniques, as it requires using reference points to analyze and process these points to achieve accurate target location estimation.  As shown in Fig. \ref{fig1}, reference points are chosen physical locations where signals are collected.  All reference points are represented  by $P=\{(x_i,y_i)|i=1,2,3\cdots M\}$, with $M$ denoting the total number of selected reference points. The signal received at the $i$-th reference point can be represented as
\begin{equation}
y_i(n)=s(n)*h_i(n)+\omega(n),
\label{eq1}
\end{equation}
where $y_i(n)$ denotes the received baseband signal, $s(n)$ denotes the waveform generated by the FM transmitter, $h_i(n)$ denotes the impulse response of the propagation channel between the FM transmitter and the $i$-th reference point, $*$ denotes convolution, and $w(n)$ is the noise usually modeled as additive white gaussian noise (AWGN).


Due to the significant differences in communication mediums (such as air, walls, pedestrians, and iron objects) between the transmitter and receiver of FM devices in different environments, the signal received at the same location fluctuates over time, exhibiting substantial diversity in FM signals. This variability poses a significant challenge to positioning techniques. To accurately estimate the target location, we first collect FM signal data at several reference points. Based on these data, a mapping function $L$ is obtained to associate locations. During the positioning phase, this mapping is used to determine the most probable position $P^{\prime}$ of the signal received at the current sampling point, expressed as $P^{^{\prime}}=L(y_i(n))$. To achieve precise positioning, it is crucial to minimize the positioning error as much as possible:
\begin{equation}
\min_LD(P,P^{^{\prime}}),
\label{eq2}
\end{equation}
where $D(\cdot,\cdot)$ represents the distance between these positions. 

\section{Methodology}
\label{sec2}

\subsection{Overall Framework}

\begin{figure*}[tp]
    \centering
    \includegraphics[width=0.8 \textwidth]{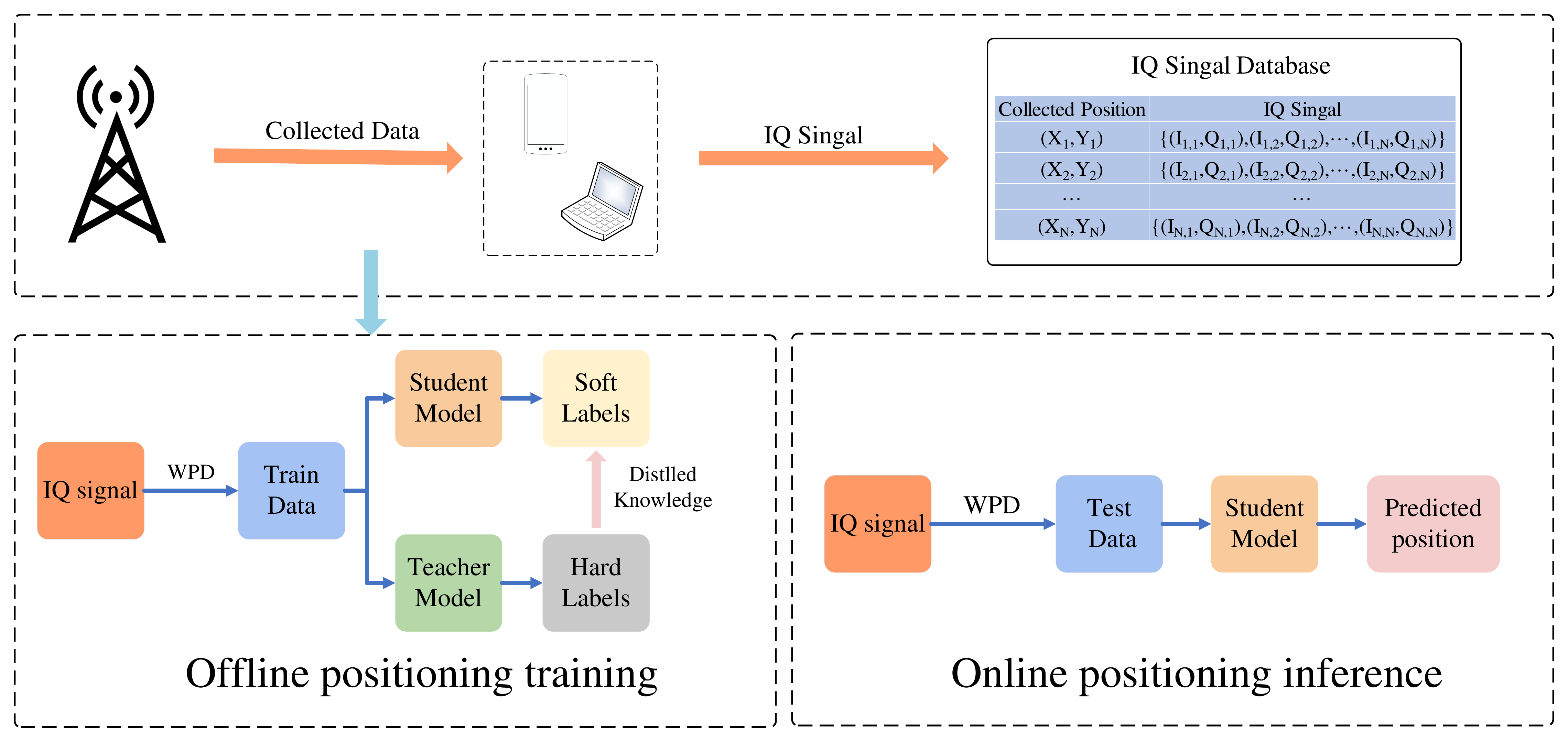}
    \caption{The overall framework of the proposed positioning method WK-Pnet.}
    \label{fig:WK-Pnet}
\end{figure*}

In this study, we propose a deep learning-based FM signal positioning system named WK-Pnet, as illustrated in Fig. \ref{fig:WK-Pnet}. The framework consists of two main stages: offline training and online inference. The core objective of this system is to use collected IQ signal data to train a deep learning model, enabling accurate target location prediction while maintaining model lightweighting.

During the offline training phase, we first build a database containing FM signals, comprising signal samples collected from various positions. These samples are used to train the teacher model, enabling it to learn localization capabilities based on FM signal features. To ensure the model's positioning accuracy and lightweighting, we apply knowledge distillation. In this process, the previously trained teacher model serves as a pre-trained model, and its output is used as a soft target to guide the training of the student model. \textcolor{blue}{The student model is trained using the same dataset as the teacher model. By learning from the teacher model’s predictions, the student model continuously optimizes its positioning capability. } Notably, we employ WPD to further extract FM signal features, enhancing the model’s understanding and representation of signal characteristics.


In the online inference phase, the trained student model processes the received FM signals to estimate the target's actual position. In this stage, the student model leverages knowledge acquired during the offline training phase to analyze new FM signal data, subsequently estimating the target’s position with a Bayesian confidence-based approach.

Overall, the WK-Pnet system combines wavelet packet decomposition, deep learning, and knowledge distillation to create an efficient and accurate wireless signal positioning framework. This system not only learns rich signal features during offline training but also enables fast and precise target location prediction during online inference.

\subsection{Data Processing}



Wavelet transforms, known for their excellent time-frequency positioning properties, have brought significant advances to the field of signal processing. They capture subtle signal variations and provide deeper insights into the intrinsic structure of signals through multi-scale analysis. In this study, we employ WPD, which offers finer frequency band divisions and can reveal multiple layers of signal characteristics. Compared to traditional wavelet analysis, WPD allows for more detailed frequency band partitioning, enabling us to select the most suitable wavelet basis based on signal characteristics and thus enhancing the accuracy and efficiency of time-frequency analysis.

WPD uses low-pass and high-pass filters to meticulously decompose the input time series, separating the signal into approximate and detailed parts. Given the orthogonal scaling function $\phi(t)$ and wavelet function $\psi(t)$ the two-scale equation of WPD can be expressed as:
\begin{equation}
\begin{cases}\phi(t)=\sqrt{2}\sum_{k\in Z}h_{0k}\phi(2t-k)\\\psi(t)=\sqrt{2}\sum_{k\in Z}h_{1k}\phi(2t-k)\end{cases}
\label{eq3}
\end{equation}
where $h_{0k}$ and $h_{1k}$ are the filter coefficients in the multi-resolution analysis and $Z$ represents an integer set.
At each decomposition step, the input signal is split into a low-frequency rough approximation and a high-frequency detail part. To extend the two-scale equation, the following recursive relation is defined:
\begin{equation}
\\\begin{cases}\ w_{2n}(t)\ \ =\sqrt{2}\sum_{k\in Z}h_{0k}w_n(2t-k)\\w_{2n+1}(t)=\sqrt{2}\sum_{k\in Z}h_{1k}w_n(2t-k)\end{cases}
\label{eq4}
\end{equation}
where: when $n=0$, $w_0=\phi(t)$, $w_1=\psi(t)$ ; $\{w_n(t)\}$ is a set of basis functions determined by the wavelet $w_n=\phi(t)$.

\begin{figure}[t]
    \centering 
    \includegraphics[width=7.5cm]{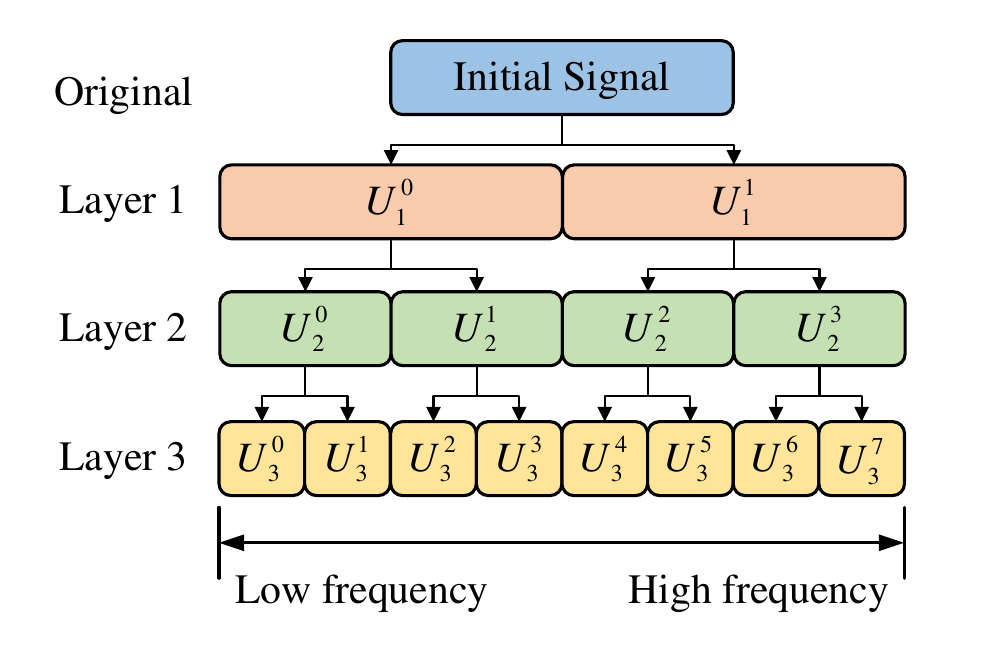}  
    \caption{Tree structure of WPD.}
    \label{fig3}
\end{figure}

Fig. \ref{fig3} shows the tree structure of WPD, where each node represents the decomposition of the signal at a specific frequency and time scale. The root node of the tree corresponds to the approximation part of the original signal, while the child nodes represent the detail components. In WPD, each node is further divided into two child nodes: one representing the high-frequency component (detail) and the other representing the low-frequency component (approximation). This decomposition method allows us to recursively explore the multi-scale characteristics of the signal. For example, in the first decomposition, the two child nodes of the root node represent the high-frequency detail and low-frequency approximation of the signal, respectively. As the decomposition proceeds, each child node is further decomposed, resulting in finer frequency and time representations.

When receiving an FM signal $y(n)$ of length $N$, we first decompose it into its real and imaginary components, denoted by $I(n)$ and $Q(n)$, respectively. Specifically, $I(n)$ represents the in-phase (I) component of $y(n)$, while $Q(n)$ represents its quadrature (Q) component. This decomposition can be expressed as:
\begin{equation}
I(n) = \mathrm{Re}(y(n)),\quad Q(n) = \mathrm{Im}(y(n)),
\label{eq5}
\end{equation}
Next, we apply an $L$-level WPD to both the real and imaginary parts of the signal, resulting in $2^L$ sub-signals. Each of these sub-signals corresponds to a node in the decomposition tree. By rearranging these nodes, we obtain two matrices representing the real and imaginary components in the frequency domain. Finally, we combine these two matrices into a three-dimensional matrix with a depth of 2, which is then used as the model input.

\begin{figure}
\centering
\subfigure[]{
\label{fig:WPD_plot:a}
\begin{minipage}[b]{0.22\textwidth}  
\includegraphics[width=1\textwidth]{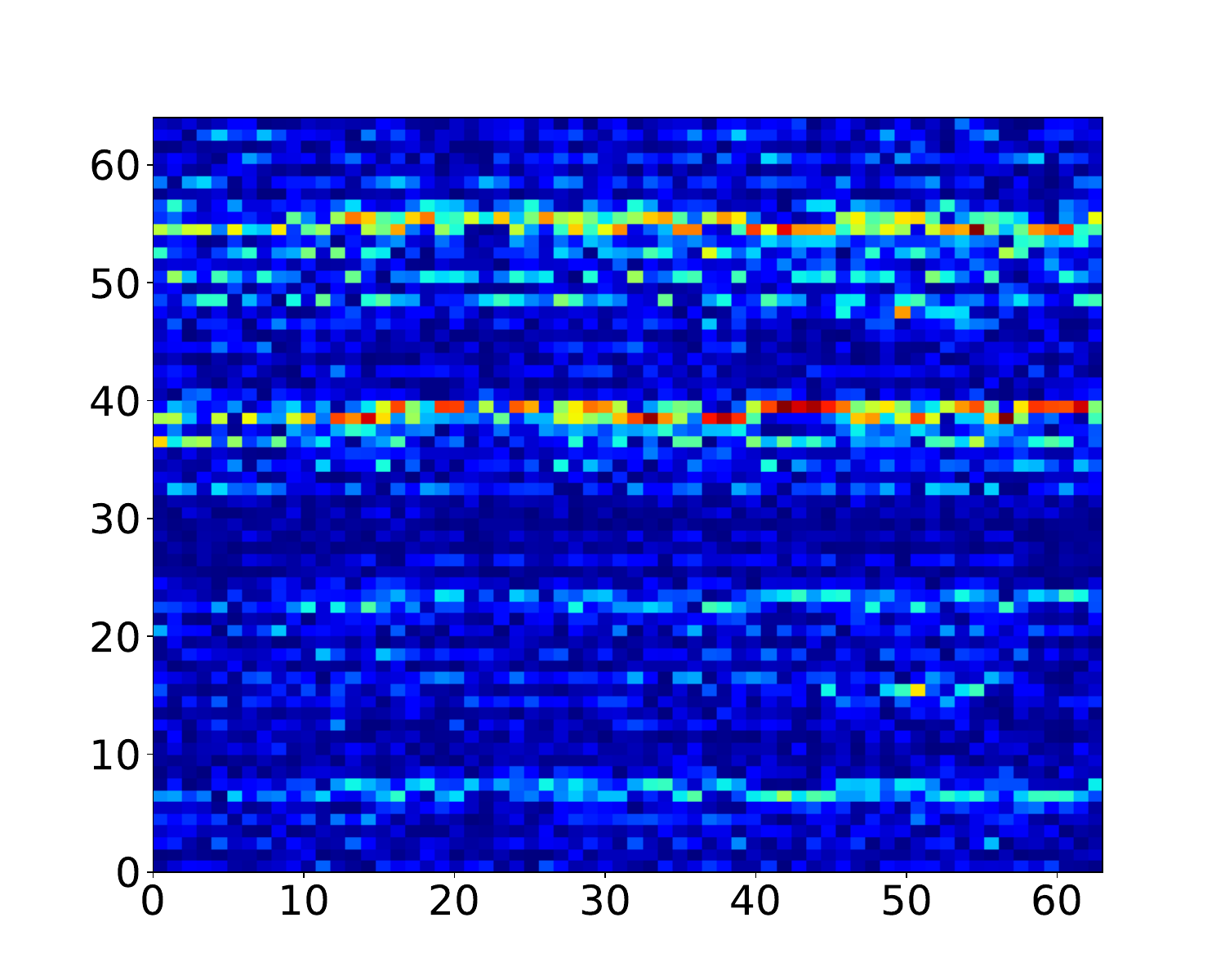} 
\end{minipage}
}
\subfigure[]{
\label{fig:WPD_plot:b}
\begin{minipage}[b]{0.22\textwidth}  
\includegraphics[width=1\textwidth]{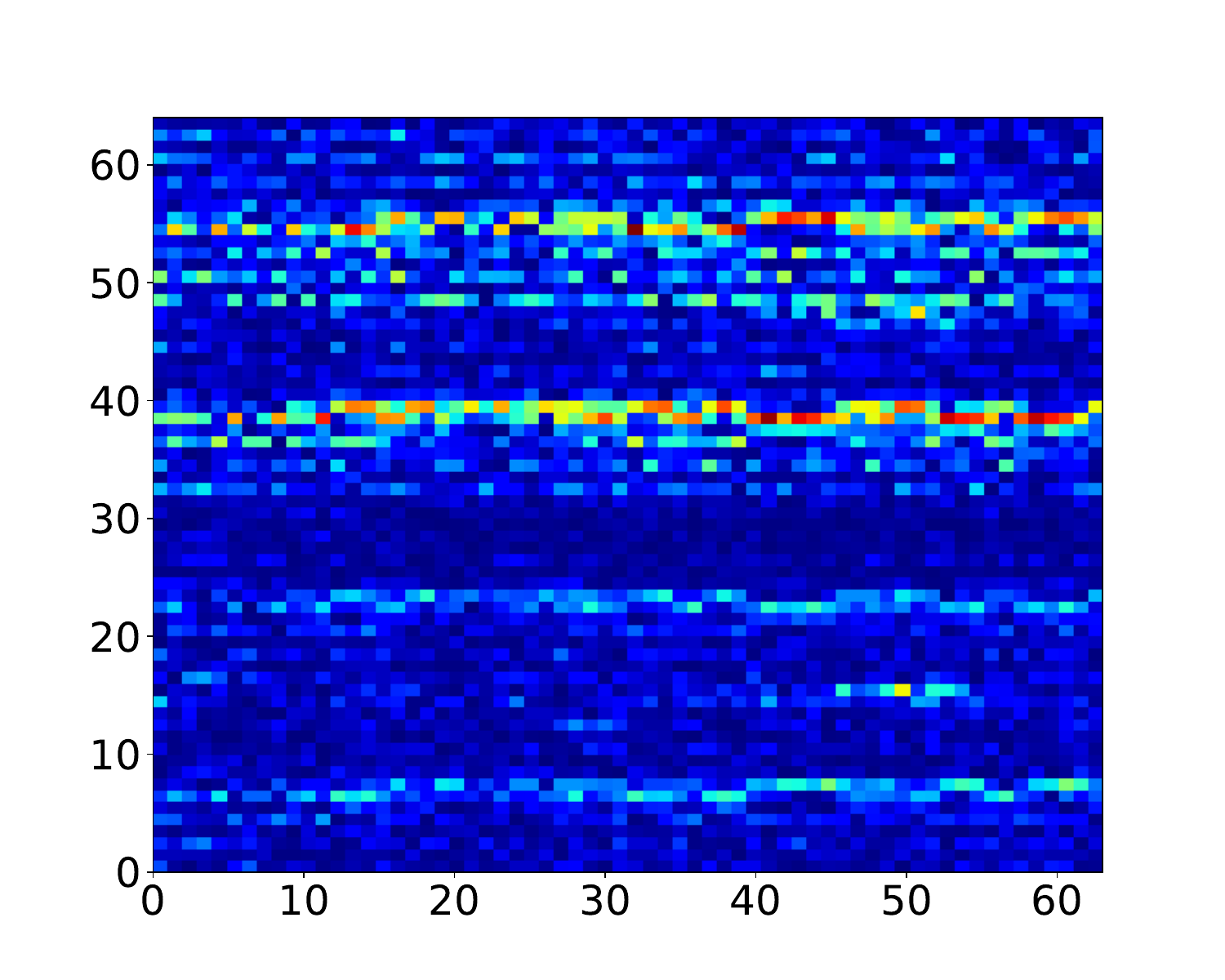} 
\end{minipage}}
\caption{The time-frequency representation of the FM signal, with the WPD level set to 6. (a) WPD of the I component, (b) WPD of the Q component.}
\label{fig:WPD_plot}
\end{figure}

Fig. \ref{fig:WPD_plot} presents the grayscale coefficient images of the root nodes after six-level WPD of the received FM signal using the Haar wavelet basis. In this figure, the horizontal axis represents the time domain of the signal, while the vertical axis corresponds to the root node coefficients obtained from the wavelet decomposition. The variation in shading across the image visually reflects the distribution of root node coefficients, with darker shading indicating higher signal intensity at that node.

\subsection{Knowledge Distillation}
The WK-Pnet method uses a knowledge distillation strategy \cite{hinton2015distilling} to transfer knowledge from a complex deep neural network (teacher model) to a simpler neural network (student model). This approach reduces model complexity, lowers computational demands, and successfully maintains model performance. Compared to traditional hard labels (labels containing only 0s and 1s), soft labels generated by the teacher model (containing probability values) can capture subtle relationships and similarities between different classes more finely. Soft labels provide the student model with richer learning materials, allowing it to gain a deeper understanding of and differentiate between categories during training, thereby enhancing its recognition and classification abilities.

\subsubsection{Teacher Model}

In the building process of the teacher model, we fine-tune the model framework based on FM-Pnet \cite{zheng2024fm}, which adopts ResNeXt \cite{xie2017aggregated} as the core network. ResNeXt combines the simplicity of the VGG design with the residual connections of ResNet while introducing grouped convolutions to enhance performance effectively. This design not only improves computational efficiency and scalability but also strengthens feature extraction capabilities through the "cardinality" module. This module consists of parallel, small sub-networks that help in learning diverse features. The structure of the teacher network is shown in Table \ref{table1}, where ``C'' represents grouped convolution with 32 groups.

\begin{table}
\caption{The Architecture of the Teacher Model}
\centering
\renewcommand\arraystretch{1.5}
\setlength{\tabcolsep}{1.5mm}{
\begin{tabular}{|c|c|c|}
\hline 
Layers & OutputSize & Detail \\
\hline 
Attention & $128 \times 32 \times 2$ &  Spatial-Attention \\ 
\hline 
conv1 & $128 \times 32 \times 64$ & $3 \times 3 $, $64$
\\ 

\hline
conv2 & 
$128 \times 32 \times 128$ &  
$\left\{\begin{array}{l}
1\times1,64 \\
3\times3,64\text{ ($C = 32$)} \\
1\times1,128
\end{array}\right\} \times 3$ 
\\ 

\hline
conv3 & 
$64 \times 16 \times 256$ &  
$\left\{\begin{array}{l}
1\times1,128 \\
3\times3,128\text{ ($C = 32$)} \\
1\times1,256
\end{array}\right\} \times 4$ 
\\ 
\hline
conv4 & 
$32 \times 8 \times 512$ &  
$\left\{\begin{array}{l}
1\times1,256 \\
3\times3,256\text{ ($C = 32$)} \\
1\times1,512
\end{array}\right\} \times 5$ 
\\ 

\hline
conv5 & 
$16 \times 4 \times 1024$ &  
$\left\{\begin{array}{l}
1\times1,512 \\
3\times3,512\text{ ($C = 32$)} \\
1\times1,1024
\end{array}\right\} \times 3$ 
\\ 

\hline
Pool & $1 \times 1 \times 1024$ & Global average pool
\\

\hline
FC & $1 \times M $ & $G$-d FC, Softmax 
\\ 

\hline
\end{tabular}}
\label{table1}
\end{table}

The spatial attention mechanism enables the model to capture time-frequency features with greater precision. Unlike traditional methods, the spatial attention mechanism can adaptively focus on key features of the signal rather than processing all features uniformly. This selective focus strategy helps the model comprehensively capture both global and local details of the signal, significantly enhancing its robustness against noise and interference. The spatial attention mechanism uses a convolutional kernel to compress the height and width of the time-frequency feature map to 1, forming a two-dimensional feature representation. This 2D representation is then processed through a $Sigmoid$ function to generate the final spatial attention map. This process ensures that the model accurately identifies and emphasizes task-critical features while suppressing less important parts, optimizing the model's performance and accuracy. The specific calculation formula is as follows:
\begin{equation}
M_s(F_s)=\sigma(f^{7\times7}([AvgPool(F_s);MaxPool(F_s)]))
\label{eq7}
\end{equation}
where ${F_s} \in {R^{C*H*W}}$ serves as the input for spatial attention, ${AvgPool}(\cdot)$ and ${MaxPool}(\cdot)$ represent global average pooling and global max pooling, respectively, and $\sigma$ denotes the $sigmoid$ function. ${M_s}$ is the spatial attention representation for the input $F$, and ${f^{7 \times 7}}$ indicates a convolution operation with a filter size of $7 \times 7$.

\subsubsection{Student model}
The student model is a small and simple network designed to enhance performance for deployment on wearable and embedded devices, while reducing power consumption and memory impact. The structure of the student neural network is detailed in Table \ref{table2}. The network consists of three convolutional blocks, each comprising a convolutional layer, a normalization layer, a ReLU activation layer, and a max-pooling layer.

\begin{table}
\caption{The Architecture of the Student Model}
\centering
\renewcommand\arraystretch{2}
\setlength{\tabcolsep}{4mm}{
\begin{tabular}{|c|c|c|}
\hline
Layers & Detail \\
\hline
Conv2D-1 & Filters-32, Kernel-3, ReLu, MaxPool(1,2) \\
\hline
Conv2D-2 & Filters-64, Kernel-3, ReLu, MaxPool(1,2) \\
\hline
Conv2D-3 & Filters-128, Kernel-3, ReLu, MaxPool(1,2) \\
\hline
Dense &  AdaptiveAvgPool2d, Flatten  \\
\hline
\end{tabular}}
\label{table2}
\end{table}

The input data first passes through the convolutional layer to extract signal features that aid learning; it then goes through the normalization layer to reduce the risk of overfitting. Next, the ReLU activation layer introduces nonlinearity, allowing the network to learn more complex features and mitigating the vanishing gradient problem. After that, the max-pooling layer reduces the dimensionality and retains critical information, improving efficiency and effectiveness in subsequent steps. Finally, a global adaptive average pooling layer combined with a fully connected layer outputs the classification confidence probability for each position.

\subsubsection{Loss Function}
Knowledge distillation modifies the Softmax formula to adjust the probability distribution of target classes, expressed as
\begin{equation}
q_i=\frac{\exp(z_i/T)}{\sum_j\exp(z_j/T)},
\label{eq8}
\end{equation}
where $q_i $ represents the softened probability for each class output by the model, \textcolor{blue}{$z_i$ is the linear output value of the model for the $i-$th class,} and is $T$ the temperature parameter.


The student model is designed to learn both the direct information from hard labels and the rich information contained in soft labels produced by the teacher model. The loss function typically consists of two parts: the $KL$ divergence loss for soft labels and the cross-entropy (CE) loss for hard labels, as shown below:
\textcolor{blue}{
\begin{equation}
L_{KL} = \mathrm{KL}(q_i^{teacher} \mid q_i^{student}),
\label{eq9}
\end{equation}
where $q_i^{teacher}$ represents the softened probability of the $i-$th class output by the teacher model, and $q_i^{student}$ represents the softened probability of the $i-$th class output by the student model.
\begin{equation}
L_{CE} = -\sum h_i \log(q_i^1)
\label{eq10}
\end{equation}
where $h_i$ is the one-hot encoding of the true label, and $q_i^{1}$ represents the probability of the $i-$th class output by the student model when $T=1$. }

To enable the student model to learn more effectively from the teacher model, we calculate the weighted sum of these two losses to obtain the total loss function of the student model. The formula for the weighted sum is expressed as
\begin{equation}
L_{total}=\alpha\cdot L_{KL}+(1-\alpha)\cdot L_{CE},
\label{eq11}
\end{equation}
where $\alpha$ is a hyperparameter between 0 and 1, used to balance the importance of KL divergence loss and cross-entropy loss in the total loss.

\subsection{Confidence-Based Position Estimation}
The WK-Pnet system operates in two main stages: offline training and online inference. Following the completion of offline training, a distinct strategy is employed for the inference phase. As probabilistic methods generally outperform deterministic ones, we utilize a probabilistic model based on Bayes' theorem to estimate the target position. The target environment is divided into multiple equal-sized grids, with the Bayesian criterion applied to calculate the probability density at each grid point. The final position estimate for each target is then determined based on this probability density, as follows:
\begin{equation}
p(i)=\frac{e^{z_i}}{\sum_{j=1}^Ne^{z_j}}
\label{eq12}
\end{equation}
\begin{equation}
(\hat{x},\hat{y})=\sum_{i=1}^Np(i)\cdot(x_i,y_i)
\label{eq13}
\end{equation}
where $z_i$ represents the \textcolor{blue}{value} from the linear output of the WK-Pnet model, $N$ denotes the total number of marked positions divided into grids, $p(i)$ is the confidence level for the $i$-th predicted position, $(x_i,y_i)$ are the horizontal and vertical coordinates of the grid position, and $(\hat{x},\hat{y})$ are the predicted horizontal and vertical coordinates.

\section{Experiment}
\label{sec3}

To comprehensively evaluate the performance of our proposed method compared to the method introduced in \cite{zheng2024fm}, we adopt three key metrics: mean distance error (MDE), standard deviation (STD), and cumulative distribution function (CDF). These metrics reflect the positioning accuracy and stability of the model from different perspectives. MDE indicates the average Euclidean distance between the estimated and actual positions; the smaller the value, the higher the positioning accuracy. STD measures the dispersion of positioning errors, where a smaller STD indicates better stability of the positioning results. CDF provides the proportion of samples successfully positioned within a specific error range, and the CDF curve allows for an intuitive observation of performance across various error thresholds. 
\textcolor{blue}{Furthermore, to gain a more comprehensive understanding of our method’s inference performance on mobile edge devices, we conduct an in-depth analysis of its computational complexity, including FLOPs, Params, and inference latency. Note that to enhance the accuracy of efficiency evaluation, we perform real-device testing on a smartphone to obtain the inference latency.}

This comprehensive evaluation enables us to compare the performance of the two methods thoroughly, verifying the effectiveness and potential superiority of our model.In the paper, for simplicity, when referring to MDE, STD, and distance errors in the CDF, all values are in meters, and we will omit the unit, providing only numerical values. The following subsections will provide detailed numerical values and analysis of these evaluation metrics.

\subsection{Experimental Setup}
\subsubsection{Datasets}

To comprehensively assess the performance of the WK-Pnet system, we utilize the dataset from \cite{zheng2024fm}. This dataset, first released in 2024, encompasses both indoor and outdoor environments, with data collected over three consecutive days from various fixed locations. The FM signal is collected with a center frequency of 97.5 MHz, a bandwidth of 4 MHz, and a sampling rate of 5 Msps. This dataset provides diverse environmental conditions that are essential for evaluating the generalization ability and robustness of the WK-Pnet system in varied urban scenarios.

\subsubsection{Operating Environment}

The experiments are conducted on a computer featuring an AMD Ryzen 9 7945HX at 2.50 GHz and an NVIDIA GeForce RTX 4060 GPU. The model is trained using the PyTorch framework, with parameter updates handled by the AdamW algorithm. \textcolor{blue}{Knowledge distillation is applied with the $T$ set to 5 and $\alpha$ set to 0.5.} The batch size is set to 10, with a total of 20 training epochs. The initial learning rate is 0.001, and it is halved every two training epochs. \textcolor{blue}{During edge device deployment, we use the iQOO Neo5 smartphone, equipped with a Qualcomm Snapdragon 870 processor, as the test device to evaluate model latency.}


\subsection{The Impact of Time-Frequency Transformation}
\textcolor{blue}{
In this experiment, we conduct an in-depth analysis of the impact of time-frequency transformation techniques on positioning accuracy. We first modify the input of FM-Pnet \cite{zheng2024fm} to compare the effect of different time-frequency inputs on positioning performance. In the calculation of STFT, we use a 512-point FFT with a $75\%$ overlap and a Hanning window. Then, the real and imaginary parts of the STFT-transformed signal are extracted and concatenated into a dual-channel matrix. For EMD, we set the maximum decomposition order of the intrinsic mode functions to 6 and decompose the I and Q channel sequences separately. Afterward, the decomposed I and Q channel data are concatenated to form a dual-channel matrix. For WPD, we choose the Haar wavelet basis and perform a 5-level decomposition.}

\begin{figure}[t]
    \centering 
    \includegraphics[width=8.5cm]{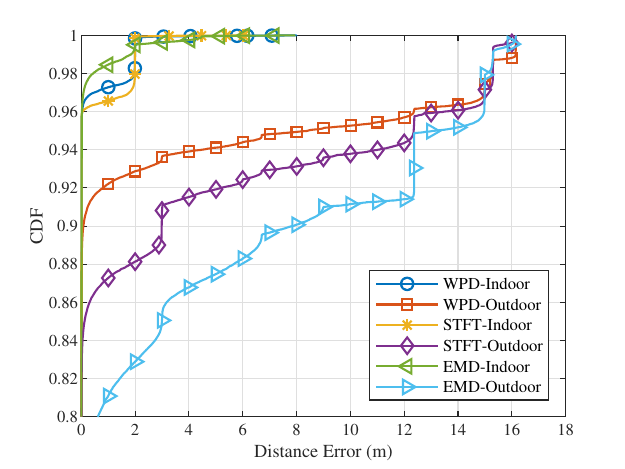}  
    \caption{\textcolor{blue}{The impact of different time-frequency transformation inputs.}}
    \label{fig5}
\end{figure}

\textcolor{blue}{
As shown in Table \ref{table3}, in the indoor environment, the MDE for STFT, WPD, and EMD are 0.0689, 0.0577, and 0.0424, respectively, with corresponding STDs of 0.3595, 0.3369, and 0.3299. In this environment, STFT has the worst positioning accuracy, with the largest error fluctuation, while EMD demonstrates more stable localization results with smaller fluctuations. In the outdoor environment, the MDE for STFT, WPD, and EMD are 1.1522, 0.8324, and 1.6840, respectively, with corresponding STDs of 3.5516, 3.2130, and 4.1041. In this case, EMD performs the worst in terms of positioning accuracy and exhibits the greatest fluctuation, while WPD demonstrates more stable positioning performance.
As shown in Fig. \ref{fig5}, in the indoor environment, the CDF curves for different time-frequency inputs are relatively close. However, in the outdoor environment, WPD converges significantly faster than STFT and EMD, indicating that WPD offers superior positioning stability and accuracy in outdoor environments. Therefore, WPD shows promising potential for application in complex environments.
}

\begin{table}[]
\caption{The impact of different time-frequency transformation inputs}
\centering
\setlength{\tabcolsep}{3.5mm}{
\begin{tabular}{|c|c|c|c|c|}
\hline
Time-Frequency      & \multicolumn{2}{c|}{Indoor}       & \multicolumn{2}{c|}{Outdoor} \\ \cline{2-5}
Transformation      & MDE       & STD       & MDE       &STD    \\ \hline
STFT                & 0.0689    & 0.3595    & 1.1522    & 3.5516 \\ \hline
WPD                 & 0.0577    & 0.3369    & 0.8324    & 3.2130 \\ \hline
\textcolor{blue}{EMD}                 & \textcolor{blue}{0.0424}    & \textcolor{blue}{0.3299}    & \textcolor{blue}{1.6840}    & \textcolor{blue}{4.1041} \\ \hline
\end{tabular}}
\label{table3}
\end{table}

\subsection{The Impact of Different Neural Network Models}
The previous experiment has demonstrated the advantage of WPD's time-frequency representation in terms of positioning performance. In this section, we analyze the impact of carefully designed neural network models on positioning performance when using WPD as the time-frequency transformation. Specifically, we compare the performance of the Teacher Model and the network model used in FM-Pnet (referred to as the Original Model). 

\begin{figure}[t]
    \centering 
    \includegraphics[width=8.5cm]{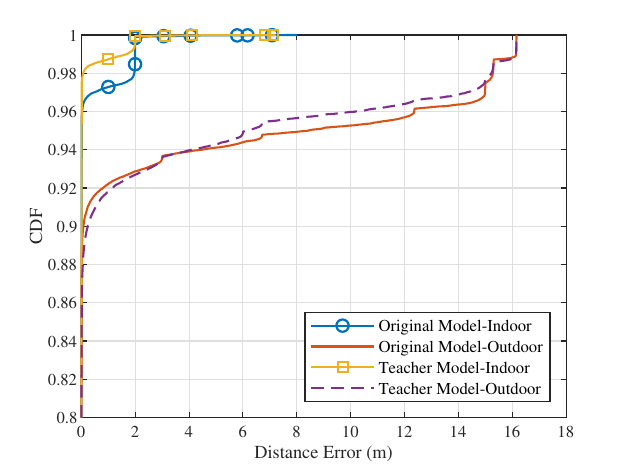}  
    \caption{The impact of different neural network models.}
    \label{fig6}
\end{figure}

As shown in Table \ref{table4}, there are significant differences in performance between different models in indoor and outdoor environments. The MDE of the Teacher Model in the indoor environment is 0.0269, compared to 0.0577 for the Original Model. This indicates that the Teacher Model has higher positioning accuracy indoors than the Original Model. The STD for the Teacher Model is 0.2252, lower than the Original Model's 0.3369, indicating that the positioning results of the Teacher Model are more stable and less volatile indoors. In the outdoor environment, the Teacher Model’s MDE is 0.7829, slightly lower than the Original Model’s MDE of 0.8324. Although the difference in MDE is small, the Original Model performs slightly worse. The Teacher Model’s STD is 3.0321, lower than the Original Model's 3.2130, demonstrating that the Teacher Model also achieves more stable positioning results in outdoor environments. This result verifies that the Teacher Model has good adaptability and robustness even in complex and variable outdoor environments.
Fig. \ref{fig6} further illustrates the distribution of positioning accuracy for different models. From the CDF curve, it is evident that the Teacher Model outperforms the Original Model in both indoor and outdoor environments. 



\begin{table}[]
\caption{Comparison of Teacher Model and Original Model}
\centering
\setlength{\tabcolsep}{3.5mm}{
\begin{tabular}{|c|c|c|c|c|}
\hline
\multirow{2}{*}{Model}   & \multicolumn{2}{c|}{Indoor}       & \multicolumn{2}{c|}{Outdoor} \\ \cline{2-5}
                    & MDE       & STD       & MDE       &STD    \\ \hline
Teacher Model       & 0.0269    & 0.2252    & 0.7829    & 3.0321 \\ \hline
Original Model      & 0.0577    & 0.3369    & 0.8324    & 3.2130 \\ \hline
\end{tabular}}
\label{table4}
\end{table}

\subsection{The Impact of WPD on WK-Pnet}
\subsubsection{The Impact of Decomposition Levels on WK-Pnet}

\begin{figure}
\centering
\subfigure[]{
\label{fig7:a}
\begin{minipage}[b]{0.4\textwidth}  
\includegraphics[width=1\textwidth]{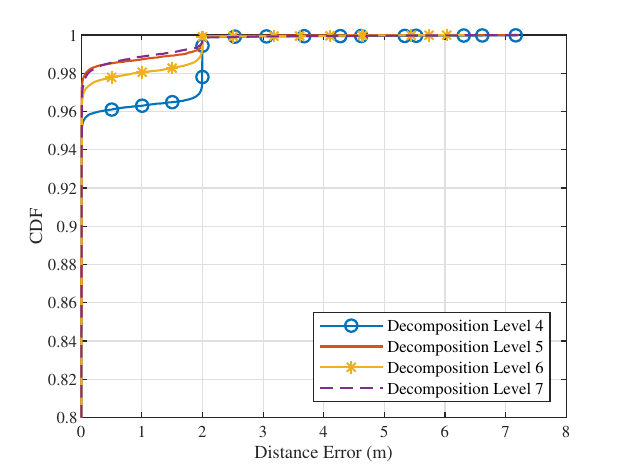} 
\end{minipage}
}
\subfigure[]{
\label{fig7:b}
\begin{minipage}[b]{0.4\textwidth}  
\includegraphics[width=1\textwidth]{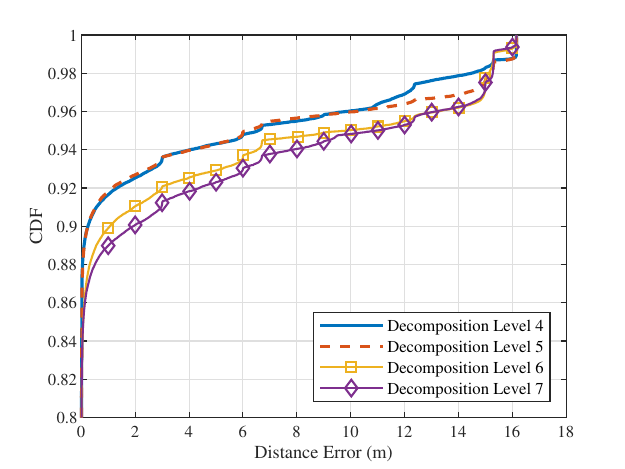} 
\end{minipage}}
\caption{The impact of decomposition levels on WK-Pnet. (a) Indoor, (b) outdoor.}
\label{fig7}
\end{figure}

First, we investigate the effect of different decomposition levels on positioning performance, selecting the Haar wavelet basis and setting the decomposition levels to 4, 5, 6, and 7. As shown in Table \ref{table5}, in indoor environments, increasing the decomposition level from 4 to 6 generally reduces the MDE, reaching a minimum of 0.0269 at levels 5 and 7. The STD also improves, with the lowest STD of 0.2252 observed at decomposition level 5, indicating greater consistency in positioning accuracy. In contrast, for outdoor environments, the MDE increases as the decomposition level rises, with decomposition level 4 yielding the lowest MDE of 0.7571, suggesting it is more suitable for outdoor settings than higher decomposition levels. In Fig. \ref{fig7:a}, decomposition levels 5 and 7 converge near the highest accuracy range for indoor positioning, demonstrating better performance in these conditions. Meanwhile, Fig. \ref{fig7:b} shows that decomposition level 4 performs best in outdoor settings, with a sharper CDF curve rise, achieving high accuracy at lower distance errors. Higher decomposition levels, such as 6 and 7, show comparatively poorer outdoor performance.

These results in both indoor and outdoor scenarios indicate that the choice of wavelet decomposition level has a significant impact on measurement error; the performance of WK-Pnet does not improve with increased decomposition levels. Choosing an appropriate decomposition level can achieve more ideal positioning results, and thus we use a decomposition level of 5 in subsequent experiments.


\begin{table}[]
\caption{Test Results of WK-Pnet with Different Decomposition Levels}
\centering
\setlength{\tabcolsep}{3.6mm}{
\begin{tabular}{|c|c|c|c|c|}
\hline
Decomposition     & \multicolumn{2}{c|}{Indoor}       & \multicolumn{2}{c|}{Outdoor} \\ \cline{2-5}
Level               & MDE       & STD       & MDE       &STD    \\ \hline
4                   & 0.0768    & 0.3956    & 0.7571    & 2.9022 \\ \hline
5                   & 0.0269    & 0.2252    & 0.7829    & 3.0321 \\ \hline
6                   & 0.0415    & 0.2825    & 0.9574    & 3.3004 \\ \hline
7                   & 0.0269    & 0.2361    & 1.0307    & 3.3844 \\ \hline
\end{tabular}}
\label{table5}
\end{table}

\subsubsection{The Impact of Wavelet Basis on WK-Pnet}

\begin{figure}
\centering
\subfigure[]{
\label{fig8:a}
\begin{minipage}[b]{0.4\textwidth}  
\includegraphics[width=1\textwidth]{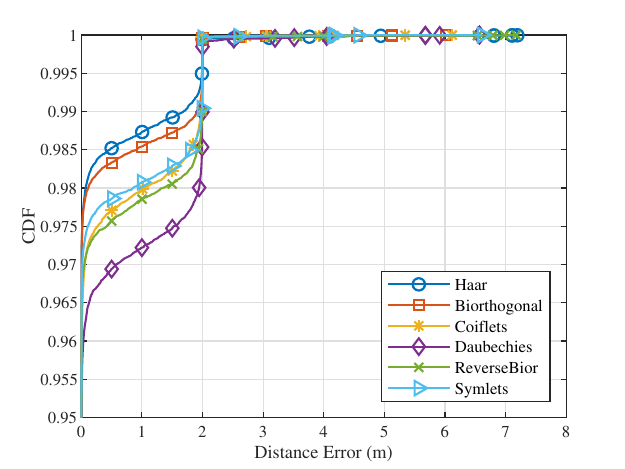} 
\end{minipage}
}
\subfigure[]{
\label{fig8:b}
\begin{minipage}[b]{0.4\textwidth}  
\includegraphics[width=1\textwidth]{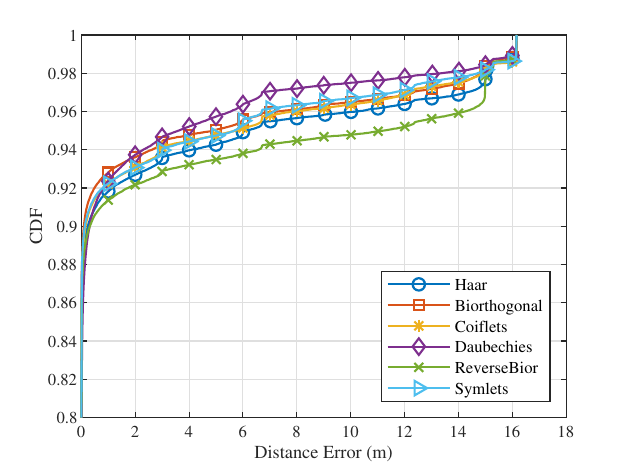} 
\end{minipage}}
\caption{The impact of wavelet basis on WK-Pnet. (a) Indoor, (b) outdoor.}
\label{fig8}
\end{figure}

Next, we analyze the effect of different wavelet bases on positioning performance. The following wavelet bases were considered in the experiment: Haar, Biorthogonal, Coiflets, Daubechies, ReverseBior, and Symlets. The experimental results are shown in Table \ref{table6} and Fig. \ref{fig8}. 

\textcolor{blue}{
As shown in Table \ref{table6}, in the indoor environment, the Haar wavelet basis has the smallest MDE and STD, which are 0.0269 and 0.2252, respectively, indicating the best positioning accuracy and stability in the indoor environment. Other wavelet bases perform relatively poorly, with the Daubechies wavelet showing the largest MDE and STD, which are 0.0573 and 0.3252, respectively, indicating the worst indoor positioning performance. The CDF curve in Fig. \ref{fig8:a} further confirms this conclusion, with the Haar wavelet base showing the fastest convergence in the indoor environment, indicating a more concentrated error distribution.}

\textcolor{blue}{
In the outdoor environment, the Daubechies wavelet base outperforms other wavelet bases, achieving the smallest MDE and STD, which are 0.5951 and 2.5047, respectively. In contrast, the ReverseBior wavelet shows relatively high errors, with MDE and STD of 0.9229 and 3.3651, respectively. This is further validated in Fig. \ref{fig8:b}, where the Daubechies wavelet base’s CDF curve converges the fastest, indicating higher positioning accuracy within a small error range. Meanwhile, the ReverseBior wavelet base’s CDF curve converges the slowest, indicating a larger overall positioning error and weaker performance.}




These results indicate that selecting an appropriate wavelet basis is crucial for improving positioning accuracy and that different wavelet bases may be more suitable for different application scenarios. Given that the Haar wavelet basis is the simplest and has the lowest computational complexity, we will use it in subsequent experiments.

\begin{table}[]
\caption{Test Results of WK-Pnet with Different Wavelet Basis}
\centering
\setlength{\tabcolsep}{3.5mm}{
\begin{tabular}{|c|c|c|c|c|}
\hline
\multirow{2}{*}{Wavelet Basis}   & \multicolumn{2}{c|}{Indoor}       & \multicolumn{2}{c|}{Outdoor} \\ \cline{2-5}
                    & MDE       & STD       & MDE       &STD    \\ \hline
Haar            & 0.0269 & 0.2252 & 0.7829 & 3.0321 \\ \hline
Biorthogonal    & 0.0301 & 0.2336 & 0.6801 & 2.8242 \\ \hline
Coiflets        & 0.0414 & 0.2697 & 0.7154 & 2.8686 \\ \hline
Daubechies      & 0.0573 & 0.3252 & 0.5951 & 2.5047 \\ \hline
ReverseBior     & 0.0444 & 0.2865 & 0.9229 & 3.3651 \\ \hline
Symlets         & 0.0390 & 0.2649 & 0.6887 & 2.7805 \\ \hline
\end{tabular}}
\label{table6}
\end{table}

\subsection{Performance of Knowledge Distillation}

In analyzing the impact of knowledge distillation, we evaluate the performance of the Teacher Model, the Student Model, and the student model without knowledge distillation (denoted as NoKD Student Model). As shown in Table \ref{table7}, in both indoor and outdoor environments, the Student Model with knowledge distillation has a lower MDE than the NoKD Student Model, indicating that knowledge distillation effectively improves the model's positioning accuracy.

Specifically, in indoor environments, the MDE of the Teacher Model is 0.0269, the MDE of the Student Model is 0.0310, and the MDE of the NoKD Student Model is 0.0464. The STD of the Teacher Model is 0.2252, the STD of the Student Model is 0.2478, and the STD of the NoKD Student Model is 0.2549. This indicates that the positioning stability of the Student Model indoors is slightly lower than that of the Teacher Model but better than that of the NoKD Student Model.

In outdoor environments, the MDE of the Teacher Model is 0.7829, the Student Model's MDE is 0.9549, and the NoKD Student Model’s MDE is 1.0918. The STD of the Teacher Model is 3.0321, the STD of the Student Model is 3.1984, and the STD of the NoKD Student Model is 3.1676. This suggests that the Student Model's positioning stability outdoors is slightly lower than that of the Teacher Model but similar to that of the NoKD Student Model.

\begin{figure}[t]
    \centering 
    \includegraphics[width=8.5cm]{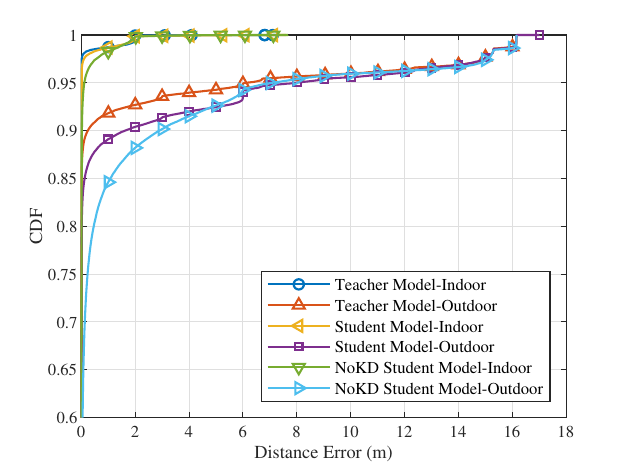}  
    \caption{Performance of knowledge distillation.}
    \label{fig9}
\end{figure}

As shown in Fig. \ref{fig9}, the CDF curves of the Student Model in both indoor and outdoor environments outperform those of the NoKD Student Model, further confirming the effectiveness of knowledge distillation in enhancing model generalization and positioning accuracy.

\begin{table}[t]
\caption{Performance of Knowledge Distillation}
\centering
\setlength{\tabcolsep}{3mm}{
\begin{tabular}{|c|c|c|c|c|}
\hline
\multirow{2}{*}{Model}   & \multicolumn{2}{c|}{Indoor}       & \multicolumn{2}{c|}{Outdoor} \\ \cline{2-5}
                & MDE    & STD    & MDE    &STD    \\ \hline
Teacher Model        & 0.0269 & 0.2252 & 0.7829 & 3.0321 \\ \hline
Student Model        & 0.0310 & 0.2478 & 0.9549 & 3.1984 \\ \hline
NoKD Student Model    & 0.0464 & 0.2549 & 1.0918 & 3.1676 \\ \hline
\end{tabular}}
\label{table7}
\end{table}

\subsection{Cross-Date Performance Analysis}

In the cross-date performance analysis of the two positioning methods, WK-Pnet and FM-Pnet, we observe their performance across different test dates (Day1, Day2, Day3) and evaluate their stability and accuracy by comparing MDE and STD in indoor environments. Note that the model used in WK-Pnet here is the Student Model. As shown in Table \ref{table8}, on Day1, both methods show good performance. WK-Pnet achieves an MDE of 0.0310 and an STD of 0.2478, while FM-Pnet's corresponding values are 0.0689 and 0.3595, indicating that both methods provide precise positioning indoors.

\begin{table}[t]
\caption{Cross-Date Test Results}
\centering
\setlength{\tabcolsep}{6mm}{
\begin{tabular}{|c|c|c|c|}
\hline
\multirow{2}{*}{Test Date}   & \multirow{2}{*}{Method}      & \multicolumn{2}{c|}{Indoor} \\ \cline{3-4}
                        &            & MDE    & STD       \\ \hline
\multirow{2}{*}{Day1}   &WK-Pnet     & 0.0310 & 0.2478 \\ \cline{2-4}
                        &FM-Pnet     & 0.0689 & 0.3595   \\ \hline
\multirow{2}{*}{Day2}   &WK-Pnet     & 2.0740 & 2.8950 \\ \cline{2-4}
                        &FM-Pnet     & 2.4397 & 2.9016   \\ \hline
\multirow{2}{*}{Day3}   &WK-Pnet     & 2.1562 & 2.4698 \\ \cline{2-4}
                        &FM-Pnet     & 2.6973 & 2.9491   \\ \hline                        
\end{tabular}}
\label{table8}
\end{table}

\begin{figure}[t]
    \centering 
    \includegraphics[width=8.5cm]{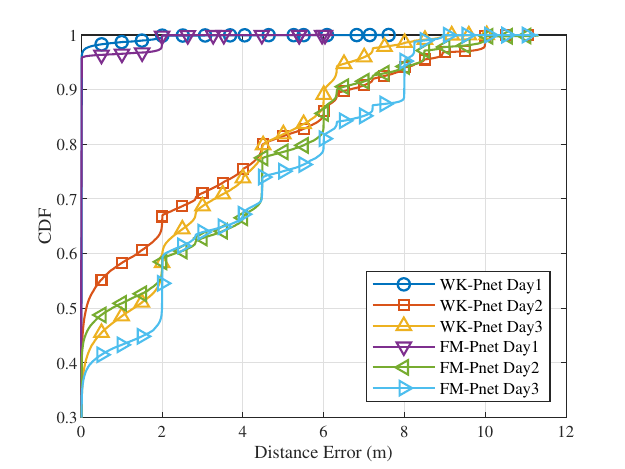}  
    \caption{Cross-date performance.}
    \label{fig10}
\end{figure}

As testing progressed to Day2 and Day3, both methods’ performance declines. On Day2, WK-Pnet's indoor MDE and STD increases significantly to 2.0740 and 2.8950, respectively, while FM-Pnet’s values are 2.4397 and 2.9016. By Day3, WK-Pnet has an MDE and STD of 2.1562 and 2.4698, respectively, while FM-Pnet's corresponding values are 2.6973 and 2.9491. Compared to FM-Pnet's results, WK-Pnet performs better in indoor environments across all dates. Additionally, As shown in Fig. \ref{fig10}, the CDF curves further demonstrate that WK-Pnet consistently outperforms FM-Pnet.

\subsection{\textcolor{blue}{Edge-end Deployment Analysis}}

\textcolor{blue}{
We conduct a detailed analysis of the FLOPs, Params, and inference latency of different time-frequency inputs and models to assess their efficiency in practical applications. As shown in Table \ref{table9}, when EMD is used as the time-frequency input, its FLOPs are the highest, reaching 9359M, while FM-Pnet’s FLOPs are 2848M. In contrast, when WPD is used as input, the Original Model's FLOPs are reduced to 1041M, demonstrating that WPD significantly reduces computational load, thus improving computational efficiency. Additionally, the Teacher Model utilizes fewer network layers. When WPD is used as the time-frequency input, its Params and FLOPs are 3.54M and 767M, respectively. Compared to the original model, the Teacher Model’s FLOPs are reduced by approximately $26.58\%$, and Params decreases by $76.35\%$. This substantial reduction makes the model more efficient during offline training and reduces the demand for computational resources and storage space.}

\textcolor{blue}{
In terms of online inference, FM-Pnet has 2848M FLOPs, while the Student Model of WK-Pnet, optimized through knowledge distillation, reduces FLOPs drastically to 117M, a reduction of about $95.9\%$. Similarly, FM-Pnet has 14M parameters, while the Student Model of WK-Pnet only has 0.10M, reducing by up to $99.3\%$.}

\textcolor{blue}{
When deployed on edge devices, different time-frequency inputs of the same model (Original Model) have a significant impact on inference latency. Among them, EMD has the highest inference latency at 219.4ms, followed by STFT at 84.5ms, while WPD achieves the fastest inference speed at only 54.5ms. Similarly, we compare the impact of different models when WPD is used as the time-frequency transformation input. After further optimizing the Original Model, the latency is reduced from 54.5ms to 49.2ms, and through knowledge distillation, it is further optimized to 8.0ms. Ultimately, WK-Pnet reduces inference latency by $90.5\%$ compared to FM-Pnet, greatly improving the model's real-time performance, making it more advantageous in resource-constrained edge computing scenarios.}

\textcolor{blue}{
These results indicate that WK-Pnet significantly reduces computational and storage requirements while ensuring sufficient positioning accuracy and greatly enhancing the speed of real-time positioning updates. This optimization makes it particularly suitable for resource-constrained environments, such as mobile devices or IoT systems, enabling more efficient and lightweight deployment while maintaining performance.
}


\begin{table}[]
\caption{\textcolor{blue}{Edge-end deployment analysis}}
\centering
 \color{blue} 
\setlength{\tabcolsep}{2.3mm}{
\begin{tabular}{|c|c|c|c|c|}
\hline

Time-frequency              & \multirow{2}{*}{Model}      & \multirow{2}{*}{\#Flops}   & \multirow{2}{*}{\#Params}      & \multirow{2}{*}{\#Latency}  \\
transformation              &                             &                            &                                &           \\ \hline
EMD                         & Original Model              & 9359M                      & 14.88M                         & 219.4ms   \\ \hline
STFT                        & Original Model              & 2848M                      & 14.88M                         & 84.5ms    \\ \hline
\multirow{3}{*}{WPD}        & Original Model              & 1041M                      & 14.88M                         & 54.5ms    \\ \cline{2-5}
                            & Teacher Model               & 767M                       & 3.54M                          & 49.2ms    \\ \cline{2-5}
                            & WK-Pnet                     & 117M                       & 0.10M                          & 8.0ms     \\ \hline

\end{tabular}}
\label{table9}
\end{table}

\section{Conclusion}
\label{sec4}

In this paper we propose WK-Pnet, a lightweight FM signal positioning method based on knowledge distillation. By leveraging WPD to extract time-frequency features from FM signals, combined with deep learning techniques, WK-Pnet achieves high-precision positioning. \textcolor{blue}{Designed to reduce computational complexity and inference latency while maintaining accuracy,} WK-Pnet is well-suited for resource-constrained environments. Experimental results demonstrate WK-Pnet’s exceptional positioning performance across various scenarios. Given its efficient, accurate positioning, WK-Pnet shows significant promise for applications requiring rapid and precise location estimation in both indoor and outdoor settings. Future research may explore combining FM signals with other signal sources to further enhance positioning accuracy and examine few-shot and self-supervised learning approaches to reduce sample requirements and labeling costs.

\bibliographystyle{IEEEtran}
\small
\bibliography{ref}

 \begin{IEEEbiography}[{\includegraphics[width=1in,height=1.25in,clip,keepaspectratio]{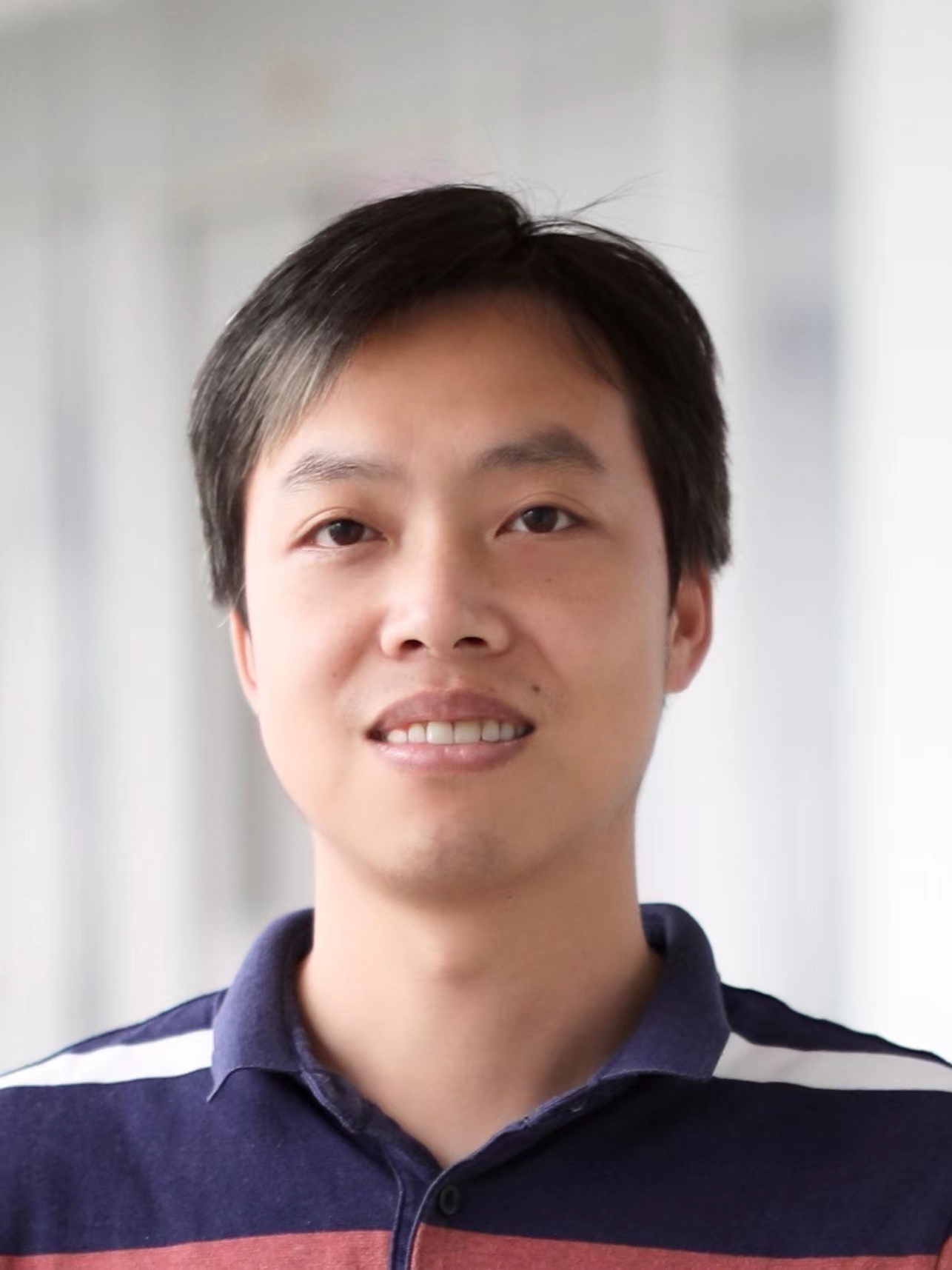}}]{Shilian Zheng}
 received the B.S. degree in telecommunication engineering and the M.S. degree in signal and information processing from Hangzhou Dianzi University, Hangzhou, China, in 2005 and 2008, respectively, and the Ph.D. degree in communication and information system from Xidian University, Xi'an, China, in 2014.

 He is currently a Researcher with National Key Laboratory of Electromagentic Space Security, Jiaxing, China, and a Doctoral Supervisor at Hangzhou Dianzi University, Hangzhou, China. His research interests include cognitive radio, deep learning-based radio signal processing, and electromagnetic space security.
 \end{IEEEbiography}

\begin{IEEEbiography}[{\includegraphics[width=1in,height=1.25in,clip,keepaspectratio]{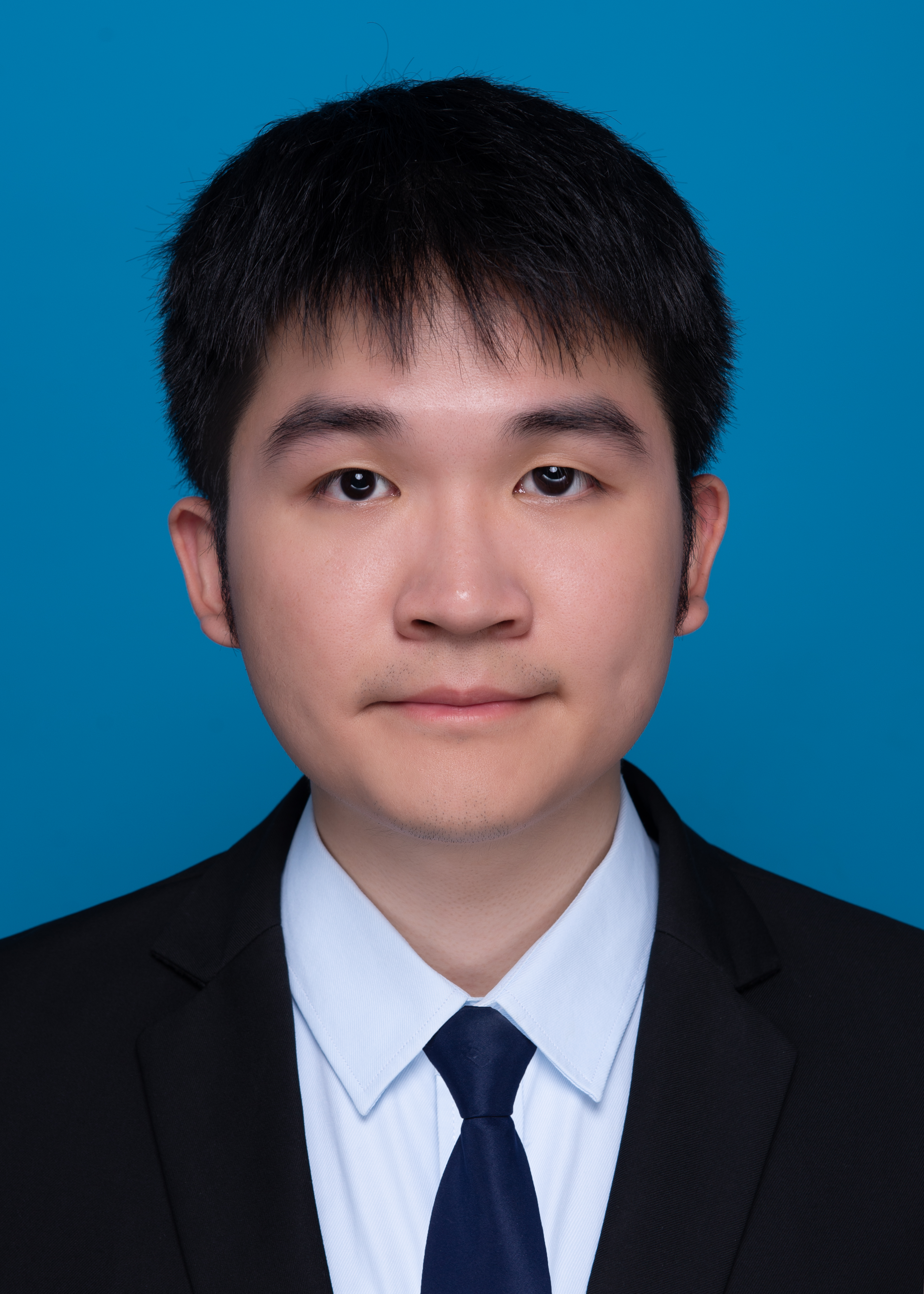}}]{Quan Lin}
received the B.S. degree in Shaoxing University, Shaoxing, China, in 2023. He is currently pursuing the M.S. degree with the School of Communications Engineering, Hangzhou Dianzi University, Hangzhou. His current research interests include deep learning and navigation via signals of opportunity.
\end{IEEEbiography}

\begin{IEEEbiography}[{\includegraphics[width=1in,height=1.25in,clip,keepaspectratio]{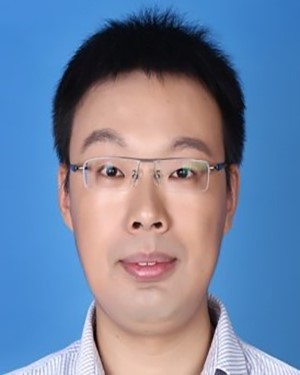}}]{Peihan Qi}
(Member, IEEE) was born in Henan, China, in 1986. He received the B.S. degree in telecommunications engineering from Chang’an University, Xi’an, China, in 2006, and the M.S. degree and the Ph.D. degree in communication and information systems from Xidian University, Xi’an, in 2011 and 2014, respectively.

He has been a Professor with the School of Telecommunications Engineering, Xidian University, since July 2022. His research interests include compressed sensing, spectrum sensing in cognitive radio networks, and intelligent signal processing. 
\end{IEEEbiography}

\begin{IEEEbiography}[{\includegraphics[width=1in,height=1.25in,clip,keepaspectratio]{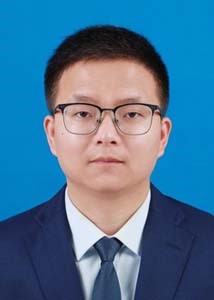}}]{Luxin Zhang}
received the M.S. degree in control science and engineering from Zhejiang University of Technology, Hangzhou, China, in 2021. 

He is currently an Assistant Engineer with National Key Laboratory of Electromagentic Space Security, Jiaxing, China. His research interests include cognitive radio, radio signal processing and learning-based radio signal recognition.
\end{IEEEbiography}

\begin{IEEEbiography}[{\includegraphics[width=1in,height=1.25in,clip,keepaspectratio]{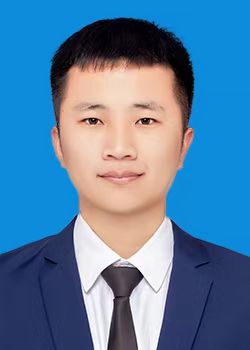}}]{Xinjiang Qiu}
received the B.S. degree in Changchun University of Science and Technology, Changchun, China, in 2023. He is currently pursuing the M.S. degree with the School of Communications Engineering, Hangzhou Dianzi University, Hangzhou. His current research interests include deep learning and navigation via signals of opportunity.
\end{IEEEbiography}




\begin{IEEEbiography}[{\includegraphics[width=1in,height=1.25in,clip,keepaspectratio]{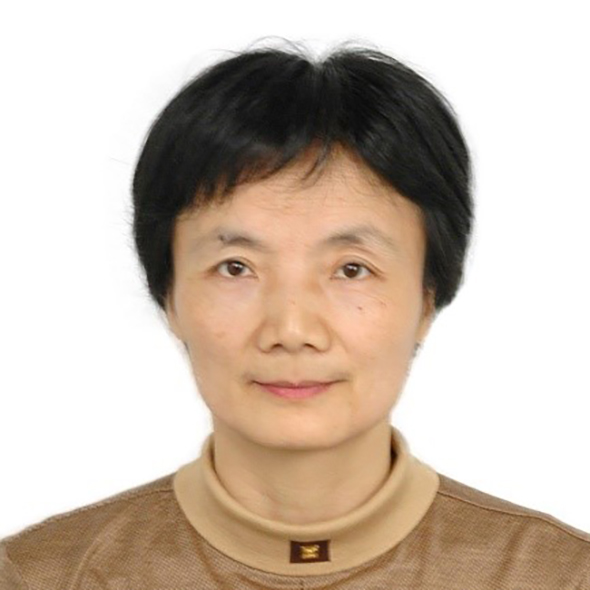}}]{Zhijin Zhao}
received the M.S. and Ph.D. degrees from from Xidian University, Xi’an, China, in 1984 and 2009, respectively. In 1993 and 2003, as a Visiting Scholar, she studied adaptive signal processing and blind signal processing in Darmstadt University of Technology and University of Erlangen-Nuremberg respectively. She is currently a Professor with the School of Communication Engineering, Hangzhou Dianzi University, Hangzhou, China. Her research interests include communication signal processing, cognitive radio technology, intelligent signal processing, and other aspects of research. She once served as President of the School of Communication Engineering, Hangzhou Dianzi University, as well as a Senior Member of China Electronics Society, a member of National Signal Processing Society.
\end{IEEEbiography}

\begin{IEEEbiography}[{\includegraphics[width=1in,height=1.25in,clip,keepaspectratio]{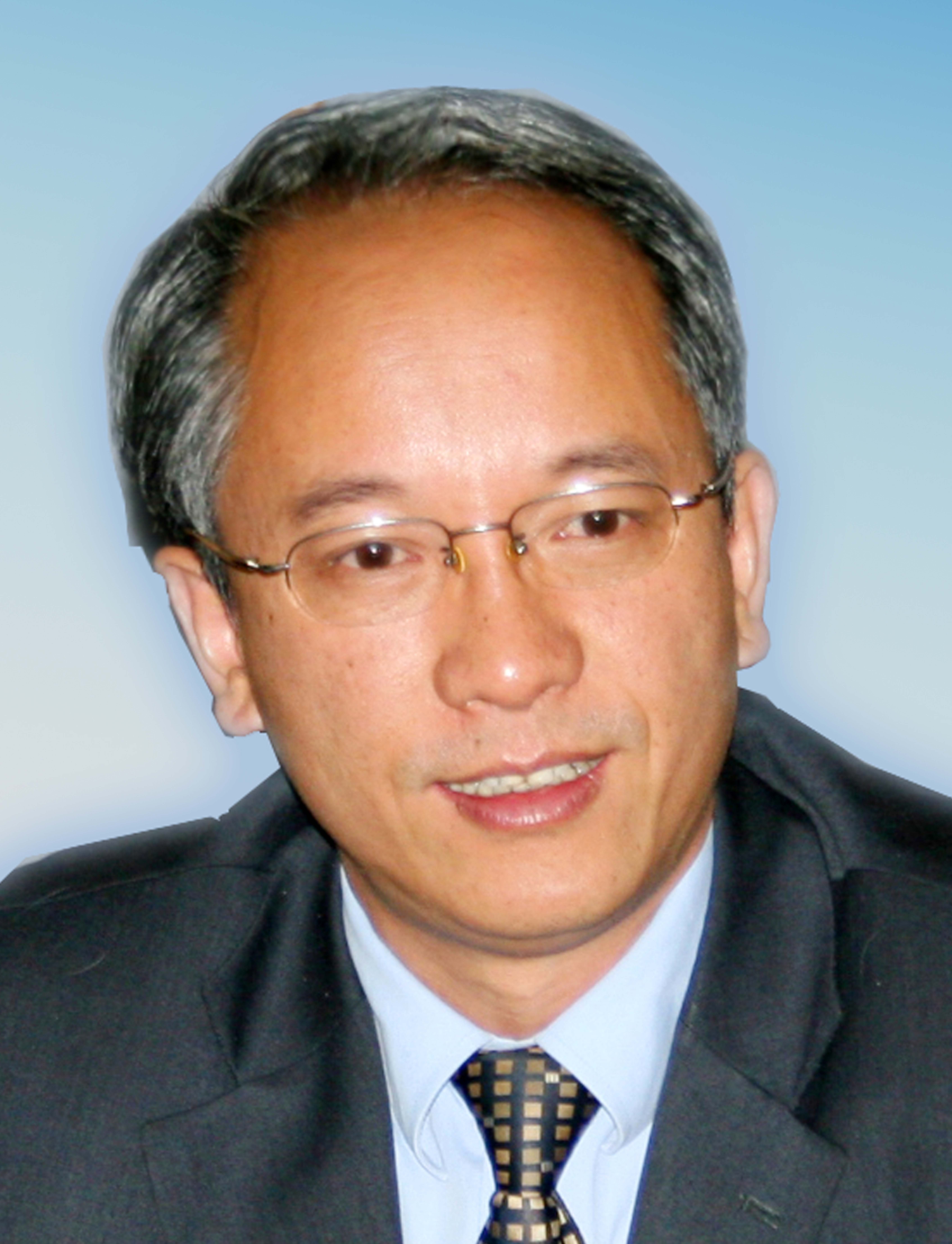}}]{Xiaoniu Yang}
is currently a chief scientist with National Key Laboratory of Electromagentic Space Security, Jiaxing, China. He is also an Academician of Chinese Academy of Engineering and a Fellow of the Chinese Institute of Electronics. He published the first software radio book in China (X. Yang, C. Lou, and J. Xu, Software Radio Principles and Applications, Publishing House of Electronics Industry, 2001 (in Chinese)). He holds more than 40 patents. His current research interests are software-defined satellite, big data for radio signals, and deep learning-based signal processing.
\end{IEEEbiography}

\end{document}